\documentclass[10pt,a4paper]{article}
\usepackage[utf8]{inputenc}
\usepackage[T1]{fontenc}
\usepackage[english]{babel}
\usepackage{newtxtext}
\usepackage{newtxmath}
\usepackage[fontsize=9.0bp]{scrextend}

\usepackage{amsmath}
\usepackage{amsfonts}
\usepackage{bm}
\usepackage{graphicx}
\usepackage[pdfa,pdfproducer={},pdfcreator={}]{hyperref}
\usepackage{url}
\usepackage{mathtools}
\mathtoolsset{showonlyrefs=true}
\usepackage{xcolor}
\usepackage{colortbl}
\definecolor{bleudefrance}{rgb}{0., 0.37, 0.67}
\hypersetup{colorlinks=true,citecolor={bleudefrance},allcolors={bleudefrance},linkbordercolor={white}}
\usepackage{microtype}
\microtypesetup{protrusion=true,expansion=true,tracking=false,spacing=true,final}
\usepackage{extdash}
\usepackage[small,compact]{titlesec}
\titleformat{\section}{\sffamily\bfseries\scshape\normalsize}{\arabic{section}}{1em}{\raggedright\MakeUppercase}
\titleformat{\subsection}{\sffamily\bfseries\small}{\arabic{section}.\arabic{subsection}}{1em}{\raggedright}
\titleformat{\subsubsection}{\sffamily\itshape\small}{\arabic{section}.\arabic{subsection}.\arabic{subsubsection}}{1em}{\raggedright}
\titlespacing{\subsubsection}{0pt}{*4}{*1}
\titlespacing{\subsection}{0pt}{*4}{*1}
\titlespacing{\section}{0pt}{*4}{*1}
\usepackage{fancyhdr}
\usepackage[left=1.350cm,right=1.350cm,top=1.5cm,bottom=2.00cm]{geometry}
\usepackage[backend=bibtex,isbn=true,url=true,eprint=false,bibstyle=numeric,sorting=none,style=numeric-comp,punctfont=false,date=year,maxbibnames=10,minbibnames=10]{biblatex}

\DeclareFieldFormat{pages}{#1}
\AtEveryBibitem{\clearfield{note}\ifentrytype{article}{\clearfield{url}\clearfield{urldate}\clearfield{issn}}{}}
\DeclareFieldFormat{doi}{\textsc{doi}\addcolon\space\ifhyperref{\href{https://doi.org/#1}{#1}}{#1}}
\DeclareCiteCommand{\citenum}{}{\printfield{labelnumber}}{}{}
\bibliography{bibs}
\usepackage{ragged2e}
\usepackage{booktabs}
\usepackage{tabularx}

\usepackage{placeins}
\newcommand{\Letter}{\textsuperscript{*}}
\usepackage{float}
\restylefloat{table}
\usepackage{caption}
\AddToHook{env/tabular/begin}{\footnotesize}

\usepackage{abstract}
\usepackage{orcidlink}
\usepackage{multicol}

\newcommand{\NIG}{\mathrm{NIG}}

\newcommand{\Nrm}{\mathcal{N}}

\begin{document}
	
	\title{\Large\textsf{\textbf{Chromatographic Peak Shape beyond the High-Péclet Approximation: Analytic Time-Domain Expression and Derivatives}}}
	\author{\small \textsf{{Hern\'an R. S\'anchez}}{\scriptsize\Letter}\orcidlink{0000-0003-1058-3396}}
	\date{}
	\maketitle
	
	\begin{center}
		{\vspace{-0.5cm}\small Instituto de F\'isica de L\'iquidos y Sistemas Biol\'ogicos, UNLP-CONICET, La Plata, 1900, Argentina
			
			\footnotesize {\scriptsize\Letter} \textsf{\href{mailto:hernan.sanchez@quimica.unlp.edu.ar}{hernan.sanchez@quimica.unlp.edu.ar}}}
	\end{center}
	
	\vspace{-0.3cm}

	\begin{abstract}
Analytic time-domain expressions for the elution-time density and its parameter derivatives were recently derived from a stochastic model of linear isocratic chromatography incorporating advective–diffusive transport, injection dispersion, and fast and slow retention mechanisms. A Gaussian approximation, justified asymptotically at high column-scale Péclet numbers, was adopted for the distribution of the combined mobile-phase and fast-retention time, while slow retention was incorporated through a series of Gaussian–Gamma convolutions.
Here, the underlying normal–inverse–Gaussian density is approximated by a finite Gaussian mixture using an existing quadrature rule. 
The elution-time PDF is a weighted sum of series of the previously derived Gaussian–Gamma form. Its parameter derivatives are also obtained.
Newly derived recurrences reduce the cost of evaluating the coefficients of that series and their derivatives from quadratic to linear in truncation order for a fixed number of mechanisms. Through a probabilistic analysis of the series coefficients, the required number of terms is found to be small enough for rapid evaluation under usual chromatographic conditions. 
A relative roundoff-error estimator is derived to quantify finite-precision loss in the inherited forward recurrence for Gaussian–Gamma convolutions. For the cancellation-prone regime, a backward reconstruction is developed, initialized from an exact ratio of consecutive Gaussian–Gamma terms expressed through parabolic-cylinder functions. A highly accurate approximation is derived for rapid evaluation of this ratio. Independent high-precision calculations confirm the high accuracy of the resulting double-precision evaluation in the tested numerical case.
	\end{abstract}
	
	\begin{multicols}{2}
		\setlength{\parindent}{1.0em}
		
		\section{Introduction}
		A mechanistic account of chromatographic peak shapes requires a representation that relates the observable profile to the physicochemical parameters associated with transport and retention processes. For this purpose, a particularly direct route is obtained in stochastic formulations, in which the contributions to the single-particle elution time from mobile-phase transport and from stationary-phase retention are treated as random variables. Under the usual linear-detection assumptions, the analyte-specific contribution to the chromatographic signal is represented, up to a scale factor, by the probability density of this elution time, commonly approximated by the intracolumn residence-time density. The physical parameters therefore enter the peak representation through the probability distributions from which this density is constructed.
		
		The earliest stochastic treatments of chromatographic retention were formulated directly in the time domain~\cite{EyringGiddings1955,GiddingsCalvin1957}, but extensions to heterogeneous surfaces in the time domain were soon held to produce expressions impractical for calculation~\cite{felinger2004}. A methodological shift followed with the systematic use of characteristic-function representations~\cite{Dondi1986,CavazziniRemelliDondiFrancesco1997, Cavazzini1999,felinger1999stochastic,pasti2016}, in which convolution of the probability densities of independent elution-time contributions corresponds to multiplication of their characteristic functions. Substantially richer models thereby became tractable, at the cost of requiring numerical inversion to recover the time-domain peak.
		The canonical stochastic picture of chromatographic peak formation combines advective--diffusive mobile-phase transport, a distributed injection profile with finite variance, and stationary-phase retention represented by one or more mechanisms, each characterized by a Poisson-distributed number of retention events
		and exponentially distributed residence times for the individual events~\cite{EyringGiddings1955, GiddingsCalvin1957, McQuarrie1963, Weiss1970, Dondi1986, CavazziniRemelliDondiFrancesco1997, Cavazzini1999, felinger1999stochastic, felinger2008, pasti2016}.
		Although these elements had long been used to represent transport, injection, and retention in stochastic chromatographic models, their
		combination into a time-domain expression fast enough for routine peak fitting remained unavailable until recently.
		
		A computationally efficient analytic time-domain expression for this stochastic--diffusive construction was derived in Ref.~\cite{Sanchez2026}. The formulation combined advection--diffusion in the mobile phase, finite initial spatial variance, and two retention mechanisms with exponentially distributed residence times: one treated as a fast contribution, associated with frequent short residence episodes, and one retained explicitly, associated with
		infrequent long residence episodes. The explicitly retained	contribution was evaluated as a series indexed by its retention-event count. This representation is efficient in the low-count regime,
		where only a small number of terms contribute appreciably, and it is physically important because a small number of long residence episodes can have a large effect on peak shape.
		The fast contribution, by contrast, has a large expected event count under typical chromatographic regimes. This conclusion was supported in the same work by a new statistical interpretation of the
		height equivalent to a theoretical plate, which yielded a conservative lower bound on the expected event count in terms of the plate number and the retention factor.
		Since the variance of a Poisson count equals its expectation, an event-count series for the fast contribution would require many non-negligible terms, making the representation inefficient.
		The fast contribution was therefore	approximated by a Gaussian density via a central-limit argument for	random sums and combined with mobile-phase transport into the mobile-plus-fast contribution.
		
		A subsequent formulation~\cite{Sanchez2026b} derived the corresponding
		analytic time-domain expression for an arbitrary number of independent,
		explicitly retained slow-retention mechanisms, each with its own event rate and
		residence-time scale. This was not an arbitrary extension: fast
		retention mechanisms whose cumulative contribution is already captured
		by their conditional mean and variance can be grouped into an effective
		Gaussian contribution. In contrast, mechanisms involving a small number
		of long residence episodes can remain shape-determining and are kept
		explicitly in the slow-retention expansion. 
		That work also introduced a faster evaluation method, which avoids direct multi-index summations in the general multi-mechanism case and, in the single-slow-mechanism case of Ref.~\cite{Sanchez2026}, reduced the computational cost by two to four orders of magnitude relative to the original analytic implementation.
		Analytical derivatives of the
		elution-time density with respect to the model parameters were also
		obtained, each evaluable at a cost comparable to that of the density.

		A step that enabled the analytic time-domain expressions of Refs.~\cite{Sanchez2026,Sanchez2026b} was the representation of the mobile-plus-fast contribution by a Gaussian density obtained as an asymptotic limit of a more specific	non-Gaussian law.
		In Ref.~\cite{Sanchez2026}, this contribution was first derived in closed form for a Dirac delta injection and identified as a Normal-Inverse-Gaussian (NIG) density~\cite{Nielsen1977}.
		Although Gaussian limits of NIG densities were already known~\cite{Nielsen1977}, the standard limiting condition does not apply to this chromatographic construction.
		The Gaussian reduction was therefore established by	direct analysis of the cumulant-generating function: conditions were derived under which its higher-order terms become numerically negligible relative to the first two terms, and these conditions were shown to hold in the high column-scale Péclet regime of the chromatographic model.
		The result was subsequently extended from idealized injections to finite initial spatial variance by treating the initial-position contribution in the high-Péclet limit through Slutsky's theorem and
		the continuous mapping theorem.
		
		The Gaussian reduction is well justified in the high column-scale Péclet regime characteristic of analytical chromatography. However, restricting the time-domain expressions to this asymptotic limit leaves the mathematical representation of the canonical physical model incomplete. At lower column-scale Péclet numbers, that asymptotic justification no longer holds. 
		The formulation developed here extends the time-domain representation beyond	the high-Péclet Gaussian reduction and recovers the previous Gaussian-base slow-retention series structure in that limit.

		The central construction is designed to preserve the computational architecture of the previous Gaussian-based time-domain formulation.	In Ref.~\cite{Sanchez2026b}, the Gaussian representation of the	mobile-plus-fast term allowed the explicit slow-retention expansion to	be organized as a single-index sum whose terms involve Gaussian--Gamma convolutions. This structure enabled low-cost recurrence-based evaluation and analytical differentiation.	The present work keeps this architecture by representing the NIG mobile-plus-fast density as a finite linear combination of Gaussian densities. This is achieved through the finite quadrature rule of Choi, Du, and Song~\cite{ChoiDuSong2020}. By linearity of convolution, each Gaussian component can then be combined with the explicit slow-retention expansion using the same Gaussian--Gamma convolution formulas.
		Consequently, analytical derivatives developed for the	previous formulation remain available component by component after the finite-mixture approximation is introduced. The resulting density and its analytical derivatives can be evaluated independently at any selected set of time points, enabling subsampling and parallel evaluation across time points for computationally efficient parameter estimation.
		
		The remaining developments address the practical reliability of the	finite representation. The auxiliary Gamma-shape index that orders the slow-retention series is identified as a finite sum of P\'olya--Aeppli random variables. 
		The moments of the summation index are then obtained in closed form from this distribution, and the truncation range is related to the physical parameters of the slow mechanisms.
		A hybrid recurrence procedure is developed to extend the accurate evaluation of the Gaussian--Gamma functions to regions where roundoff errors can be amplified during forward propagation.
		A criterion based on a propagated relative-error estimate is introduced to determine when subtractive cancellation in forward evaluation of the recurrence is estimated to raise the relative error of the computed values above a prescribed tolerance. In such cases, the recurrent sequence is reconstructed from backward ratios.
        With these developments, a numerically robust and analytically differentiable time-domain representation of this model is obtained.

		\section{Time-domain series structure with a NIG base}\label{sec:retained_structure}
Following the decoupling approximation established in Ref.~\cite{Sanchez2026}, the total residence time of an analyte particle in the column is represented as the sum of two independent random variables, $T_{mf}$ and $T_s$. The first represents the joint residence-time contribution of advective--diffusive transport in the mobile phase and the fast retention mechanism. The second  represents the contribution of the explicitly retained mechanisms. 
The deviation of the particle’s entry time into the column from the mean entry time is represented by a further independent random variable $T_{\mathrm{inj}}$. Taking the mean entry time as the time origin, the elution time is
\begin{equation}\label{eq:Tdecomp}
	T = T_{mf} + T_s + T_{\mathrm{inj}}.
\end{equation}
Although the formalism is not intrinsically restricted to slow kinetics, the practical role of $T_s$ is to represent mechanisms involving a small number of long-duration events, for which the low event count keeps the explicit expansion computationally efficient while the long residence times make their effect appreciable in the peak shape. It is therefore referred to as the slow contribution.

The random variable $T_s$ is the sum of $M$ independent random variables, one per retained mechanism. For mechanism $j$, the number of retention events is $N_j\sim\operatorname{Poisson}(\Lambda_j)$. 
The residence times of the individual events are independent and exponentially distributed with rate parameter $\theta_j^{-1}$.
The mechanisms are labeled such that
\begin{equation}
	\theta_1=\min_{1\le j\le M}\theta_j .
\end{equation}
Whenever two mechanisms share the same scale, their joint contribution to $T_s$ coincides in distribution with that of a single mechanism with that same scale and with expected event count equal to the sum of the individual $\Lambda_j$.  For convenience, such mechanisms are merged in the parametrization, yielding a unique minimum scale.

As derived in Ref.~\cite{Sanchez2026b}, the density of $T_s$ satisfies
\begin{equation}\label{eq:fTs_expansion}
	f_{T_s}(t)=e^{-\Lambda}\delta(t)
	+e^{-\Lambda}\sum_{\ell\ge1}\Omega_{\ell}\,
	f_\Gamma(t;\ell,\theta_{1}),
	\qquad
	\Lambda:=\sum_{j=1}^{M}\Lambda_j .
\end{equation}
Here, $\delta(t)$ is the Dirac delta, $f_\Gamma(\cdot;\ell,\theta_{1})$ is the Gamma density with integer shape $\ell$ and scale $\theta_{1}$, and the coefficients $\Omega_{\ell}$ depend only on the retained-mechanism parameters $\{\Lambda_j,\theta_j\}_{j=1}^{M}$.

		Let $g_b$ denote the density of $T_{\mathrm{inj}}+T_{mf}$. By independence, the elution-time density is
		
		\begin{equation}\label{eq:fT_base_series}
			f_T(t)=e^{-\Lambda}\sum_{\ell=0}^{\infty}\Omega_{\ell}\,
			h_{\ell}^{(b)}(t),
		\end{equation}
		where
		\begin{equation}\label{eq:h_base_def}
			h_{0}^{(b)}(t)=g_b(t),\qquad
			h_{\ell}^{(b)}(t)=
			\bigl(g_b*f_\Gamma(\cdot;\ell,\theta_{1})\bigr)(t),
			\quad \ell\ge1.
		\end{equation}
		
		In the previous formulations~\cite{Sanchez2026,Sanchez2026b}, $g_b$ was modeled as a Gaussian density. This identification was justified in Ref.~\cite{Sanchez2026} for the high-P\'eclet regime through analysis of the cumulant-generating function of the underlying stochastic--diffusive law. For that Gaussian-base formulation,  efficient evaluation schemes for both $h_{\ell}^{(b)}$ and its analytical derivatives with respect to the physical parameters were derived in Ref.~\cite{Sanchez2026b}.

		For an idealized Dirac-delta injection, before imposing the high-P\'eclet reduction, the mobile-plus-fast density was identified in Ref.~\cite{Sanchez2026} as Normal--Inverse--Gaussian,
		\begin{equation}\label{eq:gb_NIG}
			g_b(t)=f_{\operatorname{NIG}}(t;\alpha,\beta,\mu,\delta),
		\end{equation}
		with parameters
		\begin{equation}\label{eq:NIG_phys_map}
			\alpha=\sqrt{\frac{v^{2}}{2D_{\mathrm{eff}}\sigma_f^{2}}+\beta^{2}},
			\qquad
			\beta=\frac{1+\mu_f}{\sigma_f^{2}},
			\qquad
			\delta=\frac{\sigma_f x}{\sqrt{2D_{\mathrm{eff}}}},
			\qquad
			\mu=0.
		\end{equation}
		Here $v$ is the mean linear mobile-phase velocity along the column axis, $D_{\mathrm{eff}}$ is the effective axial diffusion coefficient for the mobile phase, $x$ is the column length, and $\mu_f$ and $\sigma_f^2$ are the mean and variance parameters of the fast-retention contribution per unit mobile-phase time.

The inverse-Gaussian quadrature of Choi, Du and Song~\cite{ChoiDuSong2020} approximates the NIG density by a finite linear combination of Gaussian densities. By linearity of convolution, the convolution $h_{\ell}^{(b)}$ with a NIG base is therefore approximated by a finite sum of Gaussian--Gamma convolutions, and the evaluation and differentiation methods derived in Ref.~\cite{Sanchez2026b} remain applicable to each term.

Let
\begin{equation}\label{eq:gamma_def}
	\gamma:=\sqrt{\alpha^{2}-\beta^{2}}.
\end{equation}
Setting $\sigma:=\sqrt{\gamma\delta}$, and letting $(z_{k},\omega_{k})_{k=1}^{n}$ denote the Gauss--Hermite nodes and weights normalised for the standard normal density, the quantities needed for the quadrature are
\begin{equation}\label{eq:uk}
	u_{k}
	=
	1+\frac{z_{k}^{2}}{2\sigma^{2}}
	+\frac{z_{k}}{\sigma}
	\sqrt{1+\frac{z_{k}^{2}}{4\sigma^{2}}},
\end{equation}
\begin{equation}\label{eq:nuk_Wk_mk}
	\nu_{k}'=\frac{\delta}{\gamma}u_{k},
	\qquad
	W_{k}=\frac{2\omega_{k}}{1+u_{k}},
	\qquad
	m_{k}=\mu+\beta\nu_{k}'.
\end{equation}
Thus,
\begin{equation}\label{eq:fNIG_linear_combination}
	f_{\NIG}(t;\alpha,\beta,\mu,\delta)
	\approx
	\sum_{k=1}^{n}W_{k}\,\Nrm(t;m_{k},\nu_{k}'),
\end{equation}
and the weights satisfy
\begin{equation}
	\sum_{k=1}^{n}W_{k}=1.
\end{equation} 
For negative $z_k$, direct floating-point evaluation of
Eq.~\eqref{eq:uk} becomes increasingly affected by subtractive
cancellation as $|z_k|/\sigma$ increases. For numerical evaluation,
it is therefore convenient to rewrite Eq.~\eqref{eq:uk} in an
algebraically equivalent form that avoids this cancellation, for example,
\begin{equation}
	u_k
	=
	\exp\!\left[
	2\,\operatorname{asinh}\!\left(\frac{z_k}{2\sigma}\right)
	\right].
\end{equation}

The distribution of the entry-time offset is approximated by
$T_{\mathrm{inj}}\sim\mathcal{N}(0,\sigma_{\mathrm{inj}}^{2})$, 
independently of the intracolumn processes.
By linearity of convolution, the Gaussian entry-time density can be convolved with each mixture component separately:
\begin{equation}\label{eq:inj_conv}
	\Nrm(t;m_{k},\nu_{k}')
	\ast
	\Nrm(t;0,\sigma_{\mathrm{inj}}^{2})
	=
	\Nrm(t;m_{k},\nu_{k}),
	\qquad
	\nu_{k}:=\nu_{k}'+\sigma_{\mathrm{inj}}^{2}.
\end{equation}
Substituting the resulting finite linear combination of Gaussian densities into Eq.~\eqref{eq:h_base_def} yields
\begin{equation}\label{eq:h_NIG_linear_combination}
	h_{\ell}^{(b)}(t)
	\approx
	\sum_{k=1}^{n}W_{k}\,
	\bigl(\Nrm(\cdot;m_{k},\nu_{k})*f_\Gamma(\cdot;\ell,\theta_{1})\bigr)(t).
\end{equation}

\section{Coefficient weights as probabilities and practical truncation}
\label{sec:finite_density}

Substituting Eq.~\eqref{eq:h_NIG_linear_combination} into Eq.~\eqref{eq:fT_base_series} produces an expression with infinite sums over the Gamma-shape index~$\ell$.
For numerical evaluation the sums are truncated at a finite maximum retained index $L\in\mathbb{N}_0$.
In Ref.~\cite{Sanchez2026b}, $L$ was determined through a residual-mass criterion that exploits the unit-sum property of the normalized slow-retention Gamma-expansion coefficients, $p_\ell=e^{-\Lambda}\Omega_\ell$, and moderate values of $L$ were found in the cases examined there.
These normalized coefficients are non-negative and sum to one, and are interpreted here as probabilities from a probability mass function of a random variable $K$.
Through this probabilistic interpretation, the required magnitude of $L$, and hence the computational cost of the truncated representation, can be explained in terms of the physical parameters.

For a fixed maximum retained index \(L\), the finite representation is
\begin{equation}
	f_T^{(L,n)}(t)
	=
	e^{-\Lambda}
	\sum_{\ell=0}^{L}\Omega_\ell
	\sum_{k=1}^{n}W_k\,h_{\ell,k}(t),
	\label{eq:fT_master}
\end{equation}
where
\begin{equation}
	h_{\ell,k}(t)
	=
	\bigl(
	\Nrm(\cdot;m_k,\nu_k)
	\ast
	f_\Gamma(\cdot;\ell,\theta_1)
	\bigr)(t),
	\qquad \ell\ge1,
	\label{eq:hlk_def}
\end{equation}
and \(h_{0,k}(t)=\Nrm(t;m_k,\nu_k)\). The coefficients \(\Omega_\ell\) are those
defined in Ref.~\cite{Sanchez2026b} and satisfy
\begin{equation}
	\Omega_0=1,
	\qquad
	\Omega_\ell
	=
	\frac{1}{\ell}
	\sum_{m=1}^{\ell}
	m\,b_m\,\Omega_{\ell-m},
	\qquad
	b_m
	=
	\sum_{j=1}^{M}
	\Lambda_j(1-a_j)a_j^{m-1},
	\label{eq:Omega_rec}
\end{equation}
where \(a_j=1-\theta_1/\theta_j\in[0,1)\). Since \(b_m\ge0\) and
\(\Omega_0=1\), it follows that \(\Omega_\ell\ge0\) for all
\(\ell\).

The normalized coefficients
\begin{equation}
	p_\ell
	:=
	e^{-\Lambda}\Omega_\ell,
	\qquad \ell\in\mathbb{N}_0 ,
	\label{eq:p_ell}
\end{equation}
are the weights of the expansion of the PDF of $T_s$. With the
convention \(f_\Gamma(t;0,\theta_1)=\delta(t)\), that expansion is
\begin{equation}
	f_{T_s}(t)
	=
	\sum_{\ell=0}^{\infty}
	p_\ell\,f_\Gamma(t;\ell,\theta_1).
	\label{eq:fTs_prob_mixture}
\end{equation}
Since each \(f_\Gamma(t;\ell,\theta_1)\) and \(f_{T_s}\) integrate to one, it follows that
\begin{equation}
	\sum_{\ell=0}^{\infty}p_\ell=1.
\end{equation}
Together with $p_\ell\ge0$, this allows $\{p_\ell\}_{\ell\ge0}$ to be interpreted as a probability mass function on~$\mathbb{N}_{0}$. Let \(K\) denote an integer-valued random variable defined by
\begin{equation}
	P(K=\ell)=p_\ell,
	\qquad \ell\in\mathbb{N}_0 .
\end{equation}
Accordingly, the coefficient mass omitted by truncation at $L$ is
\[
R_L=P(K>L)=\sum_{\ell>L}p_\ell .
\]
For fixed quadrature order $n$, this is also exactly the probability mass
omitted from the time-domain density. Since each $h_{\ell,k}$ is a
normalized density and the quadrature weights satisfy
$\sum_{k=1}^{n}W_k=1$,
\begin{equation}\label{eq:RL_density_mass}
	\int_{-\infty}^{\infty}
	\left[
	f_T^{(\infty,n)}(t)-f_T^{(L,n)}(t)
	\right]\,dt
	=
	\sum_{\ell>L}p_\ell
	=
	R_L .
\end{equation}
For $M>1$, $K$ is an auxiliary Gamma-shape index rather than the total number of physical retention events.
For a single retained mechanism, $a_1=0$, and $K$ is the Poisson-distributed number of slow-retention events. More generally, the distribution of $K$ is identified through its probability-generating function,
\begin{equation}
	G_K(u)
	=
	\mathbb{E}[u^K]
	=
	\sum_{\ell=0}^{\infty}p_\ell u^\ell .
	\label{eq:GK_def}
\end{equation}
Taking the Laplace transform of Eq.~\eqref{eq:fTs_prob_mixture} gives
\begin{equation}
	\mathcal{L}_{T_s}(s)
	=
	\sum_{\ell=0}^{\infty}
	p_\ell(1+\theta_1s)^{-\ell}.
	\label{eq:LTs_from_mixture}
\end{equation}
With
\begin{equation}
	u=(1+\theta_1s)^{-1},
	\qquad
	s=\frac{1-u}{\theta_1u},
\end{equation}
the right-hand side of Eq.~\eqref{eq:LTs_from_mixture} becomes $G_K(u)$. Therefore,
\begin{equation}
	G_K(u)
	=
	\mathcal{L}_{T_s}\!\!\left(
	\frac{1-u}{\theta_1u}
	\right).
	\label{eq:G_from_Laplace}
\end{equation}
In Ref.~\cite{Sanchez2026b}, the Laplace transform of the explicit slow-retention contribution was obtained as
\begin{equation}
	\mathcal{L}_{T_s}(s)
	=
	\exp\!\left[
	-\sum_{j=1}^{M}
	\Lambda_j
	\frac{\theta_j s}{1+\theta_j s}
	\right].
	\label{eq:LTs}
\end{equation}
Substitution of $s=(1-u)/(\theta_1u)$ into Eq.~\eqref{eq:LTs} yields
\begin{equation}
	G_K(u)
	=
	\exp\!\left[
	-\sum_{j=1}^{M}
	\Lambda_j
	\frac{1-u}{1-a_ju}
	\right].
	\label{eq:GK_explicit}
\end{equation}
The factorization
\begin{equation}
	G_K(u)
	=
	\prod_{j=1}^{M}G_j(u),
	\qquad
	G_j(u)
	:=
	\exp\!\left[
	-\Lambda_j
	\frac{1-u}{1-a_ju}
	\right],
	\label{eq:G_factorization}
\end{equation}
separates the mechanism-specific terms. The form of \(G_j\) matches the standard probability-generating function of a Pólya--Aeppli random variable, with parameters \((\Lambda_j,a_j)\) in the
parametrization of Ref.~\cite{Chukova}. Therefore, $K$ is equal in distribution to a sum of independent mechanism-specific Pólya--Aeppli components,
\begin{equation}
	K
	\overset{d}{=}
	\sum_{j=1}^{M}K_j,
	\qquad
	K_j\sim\operatorname{PA}(\Lambda_j,a_j).
	\label{eq:K_PA_sum}
\end{equation}
From this characterization, a qualitative description of the index values that can carry appreciable probability mass is obtained.
The expected value and variance of $K$ may be obtained either from the standard moments of the Pólya--Aeppli distribution \cite{Chukova} or, equivalently, by differentiating
Eq.~\eqref{eq:G_factorization} at $u=1$. They are
\begin{equation}
	\mathbb{E}[K]
	=
	\sum_{j=1}^{M}
	\frac{\Lambda_j}{1-a_j}
	=
	\sum_{j=1}^{M}
	\frac{\Lambda_j\theta_j}{\theta_1},
	\label{eq:EK}
\end{equation}
and
\begin{equation}
	\operatorname{Var}(K)
	=
	\sum_{j=1}^{M}
	\frac{\Lambda_j(1+a_j)}{(1-a_j)^2}
	=
	\sum_{j=1}^{M}
	\Lambda_j
	\frac{\theta_j(2\theta_j-\theta_1)}{\theta_1^2}.
	\label{eq:VK}
\end{equation}
Since
\begin{equation}
	\mathbb{E}[T_s]
	=
	\sum_{j=1}^{M}\Lambda_j\theta_j,
\end{equation}
Eq.~\eqref{eq:EK} can be written as
\begin{equation}
	\mathbb{E}[K]
	=
	\frac{\mathbb{E}[T_s]}{\theta_1}.
	\label{eq:EK_identity}
\end{equation}
Thus $\mathbb{E}[K]$ is the expected slow-retention time expressed in units of
the smallest retained time scale.

For each component, the index of dispersion (variance-to-mean ratio) is
\begin{equation}
	\frac{\operatorname{Var}(K_j)}{\mathbb{E}[K_j]}
	=
	\frac{1+a_j}{1-a_j}
	=
	2\frac{\theta_j}{\theta_1}-1.
	\label{eq:dispersion_index}
\end{equation}
For a single retained mechanism, \(a_1=0\), so \(K\) reduces to the Poisson
event count and
\[
\mathbb{E}[K]=\operatorname{Var}(K)=\Lambda_1.
\]
For \(a_j>0\), the corresponding Pólya--Aeppli component is overdispersed, and
its dispersion increases with the ratio $\theta_j/\theta_1$.
For the full auxiliary index,
\begin{equation}
\frac{\operatorname{Var}(K)}{\mathbb{E}[K]}
=
\sum_{j=1}^{M}
w_j
\left(
2\frac{\theta_j}{\theta_1}-1
\right),
\quad
w_j
:=
\frac{\Lambda_j\theta_j}
{\sum_{k=1}^{M}\Lambda_k\theta_k}
=
\frac{\mathbb{E}[T_{s,j}]}{\mathbb{E}[T_s]} .
\label{eq:K_dispersion_index}
\end{equation}
Thus the index of dispersion of \(K\) is a weighted average of the
mechanism-specific indices of dispersion. Each weight is the fraction of the mean explicitly retained slow-retention time contributed by the corresponding
mechanism.

In the present representation, the coefficients are accumulated from $\ell=0$ up to the truncation index. Thus a large mean, a large variance, or both, imply that a large number of terms may carry appreciable probability mass.

The relation $\mathbb{E}[K]=\mathbb{E}[T_s]/\theta_1$ connects the mean auxiliary index to the physical slow-retention time scale: for a given $\mathbb{E}[T_s]$, a smaller $\theta_1$ corresponds to a larger \(\mathbb{E}[K]\). If \(\theta_1\) is very small, the mechanism that sets this scale contributes appreciably to $\mathbb{E}[T_s]$ only when $\Lambda_1\theta_1$ is non-negligible, hence
only for large $\Lambda_1$. Such a high-count mechanism is a natural candidate for inclusion in the continuous base component.

For moderate \(\Lambda_j\), and with $\theta_1$ restricted to the time scales retained explicitly as slow mechanisms, a large ratio \(\theta_j/\theta_1\) makes the
contribution of mechanism \(j\) to both the mean and the variance of \(K\)
large. In physical time, this corresponds to a large contribution
\(\Lambda_j\theta_j\) to the mean slow-retention time, which may fall outside
the experimental time window of interest. If \(\Lambda_j\) is instead very
small, only a fraction \(1-e^{-\Lambda_j}\simeq\Lambda_j\) of analyte particles
undergoes any event of that mechanism. Such rare, late events may affect the
mean, but they do not appreciably modify the mode or the bulk shape of the
peak. Thus, regimes requiring extremely large retained indices will typically lie outside the experimental window or correspond to low-probability late events, and are not the practical regimes targeted by the explicit slow-retention expansion.
These statistical and physical considerations explain why the required truncation order is expected to remain low enough for rapid evaluation under usual chromatographic conditions.

\subsection{Computation of coefficient weights}
Direct evaluation of the $\Omega_\ell$ coefficients involves combinations of
event counts across the retained mechanisms. Because the number of such
combinations grows rapidly with both the number of mechanisms and the retained
order, the computation becomes prohibitive even for moderate values of $M$ and
$L$. In Ref.~\cite{Sanchez2026b}, this combinatorial bottleneck was eliminated
by deriving the single-index recurrence Eq.~\eqref{eq:Omega_rec}, which makes
the coefficient calculation feasible and fast in practice. Below, the structure
of that recurrence is exploited to reduce its scaling with $L$ to linear,
significantly reducing the computational cost of evaluating
$\Omega_0,\ldots,\Omega_L$.

Substitution of the expression for $b_m$ into the coefficient recurrence gives
\begin{equation}
	\Omega_\ell
	=
	\frac{1}{\ell}
	\sum_{j=1}^{M}
	\Lambda_j(1-a_j)
	C_{j,\ell},
	\qquad
	C_{j,\ell}
	:=
	\sum_{m=1}^{\ell}
	m a_j^{m-1}\Omega_{\ell-m}.
	\label{eq:Omega_C_rec}
\end{equation}
The repeated evaluation of $C_{j,\ell}$ at successive orders $\ell$ is the
costly step in this coefficient evaluation and the part whose computation can
be optimized. Separating the term $m=1$ and substituting $r=m-1$ in the
remaining sum yield
\begin{equation}
	C_{j,\ell+1}
	=
	\Omega_\ell
	+
	a_j
	\sum_{r=1}^{\ell}
	(r+1)a_j^{r-1}\Omega_{\ell-r}.
	\label{eq:C_update_start}
\end{equation}
Splitting the factor $r+1$ into its $r$ and $1$ contributions separates the
remaining sum into $C_{j,\ell}$ and a second sum. Defining
\begin{equation}
	D_{j,\ell}
	:=
	\sum_{r=1}^{\ell}
	a_j^{r-1}\Omega_{\ell-r},
	\label{eq:D_def}
\end{equation}
it follows that
\begin{equation}
	C_{j,\ell+1}
	=
	\Omega_\ell
	+
	a_j C_{j,\ell}
	+
	a_j D_{j,\ell}.
	\label{eq:C_update}
\end{equation}
Applying analogous steps to $D_{j,\ell+1}$ gives
\begin{equation}
	D_{j,\ell+1}
	=
	\Omega_\ell
	+
	a_jD_{j,\ell}.
	\label{eq:D_update}
\end{equation}
Since $\Omega_0=1$, the initial values are
\begin{equation}
	C_{j,1}=1,
	\qquad
	D_{j,1}=1.
	\label{eq:CD_initial}
\end{equation}

\section{Gaussian--Gamma recurrence and hybrid propagation}\label{sec:stability}
	The evaluation of the $h_{\ell,k}(t)$ is required  to compute both the density $f_T^{(L,n)}$ and its analytical derivatives.	
	The sequence $\{h_{\ell,k}(t)\}_{\ell=0,1,2,\dots}$ admits an inexpensive three-term forward recurrence in the Gamma-shape index. Once $h_{0,k}$ and $h_{1,k}$ have been evaluated, each additional term requires only a small, fixed number of arithmetic operations.
	
	For each component $k$, define 
	\begin{equation}\label{eq:rho_kappa}
	\rho_{k}=\frac{\nu_{k}}{\theta_{1}^{2}},\qquad
	\kappa_{k}(t)=\frac{t-m_{k}}{\theta_{1}}-\frac{\nu_{k}}{\theta_{1}^{2}}.
	\end{equation}
	The sequence satisfies
	\begin{equation}\label{eq:rec_h}
	(\ell-1)h_{\ell,k}(t)=\rho_{k}h_{\ell-2,k}(t)+\kappa_{k}(t)h_{\ell-1,k}(t),
	\qquad\ell\ge2,
	\end{equation}
	with $h_{0,k}(t)=\Nrm(t;m_{k},\nu_{k})$ and $h_{1,k}(t)$ obtained from the
	corresponding expression in Ref.~\cite{Sanchez2026b} under the substitution
	$(\mu_G,\sigma_G^2)\to(m_k,\nu_k)$.

	In finite-precision evaluation of this recurrence, subtractive cancellation can, under specific conditions, reduce the relative accuracy of the computed values. 
	This local loss of precision does not necessarily imply a comparable loss in the total density or its derivatives, due to the damping effect of the expansion coefficients.
	However, it is useful to improve numerical robustness by estimating the accumulated relative roundoff error during forward propagation and using an alternative representation when the estimate exceeds a prescribed tolerance.
	
	Because $\nu_k>0$, $\theta_1>0$, and $h_{\ell,k}(t)>0$, the term
	$\rho_k h_{\ell-2,k}(t)$ is positive.  Subtractive cancellation in Eq.~\eqref{eq:rec_h} can therefore occur only when
	$\kappa_k(t)<0$. In that case, Eq.~\eqref{eq:rec_h} contains
	the difference
	\begin{equation}
		\rho_k h_{\ell-2,k}(t)
		-
		|\kappa_k(t)|h_{\ell-1,k}(t).
	\end{equation}
	The condition $\kappa_k(t)<0$ is necessary for cancellation but it does not imply that the forward step loses significant precision.
	The switching criterion is therefore based on a propagated error estimate.
	
Let $\widehat h_{\ell,k}(t)$ denote the computed value of $h_{\ell,k}(t)$,
and let $E_{\ell,k}(t)$ denote an estimate of its accumulated absolute
error. For the purpose of this indicator, the error estimates for the starting values are initialized at the unit-roundoff level,
\begin{equation}
	E_{0,k}(t)=u_{\rm r}\,|\widehat h_{0,k}(t)|,
	\qquad
	E_{1,k}(t)=u_{\rm r}\,|\widehat h_{1,k}(t)|,
	\label{eq:E_initial}
\end{equation}
where $u_{\rm r}$ is the unit roundoff.

	For $\kappa_k(t)<0$, the forward recurrence is monitored as follows. For a
	candidate forward step, define
	\begin{equation}
		\widehat h_{\ell,k}^{F}(t)
		=
		\frac{
			\rho_k\widehat h_{\ell-2,k}(t)
			+
			\kappa_k(t)\widehat h_{\ell-1,k}(t)
		}{\ell-1}.
		\label{eq:h_forward_candidate}
	\end{equation}
	The corresponding accumulated error estimate is
\begin{align}
	E_{\ell,k}^{F}(t)
	&=
	\frac{
		\rho_k E_{\ell-2,k}(t)
		+
		|\kappa_k(t)|E_{\ell-1,k}(t)
	}{\ell-1}\\
	&+
	c_{\rm op}u_{\rm r}\,
	\frac{
		\rho_k|\widehat h_{\ell-2,k}(t)|
		+
		|\kappa_k(t)|\,|\widehat h_{\ell-1,k}(t)|
	}{\ell-1}.
	\label{eq:E_forward}
\end{align}
In the first term, the error estimates associated with the two previous recurrence values are propagated. In the second term, the local roundoff introduced by the arithmetic operations of the current recurrence step is estimated. Here $c_{\rm op}$ is a factor introduced to account approximately for the rounded elementary operations in one forward step. A simple choice is $c_{\rm op}=4$, corresponding to two multiplications, one addition or subtraction, and one division.

	The relative error estimate associated with the candidate forward value is
	\begin{equation}
		r_{\ell,k}^{F}(t)
		=
		\frac{E_{\ell,k}^{F}(t)}
		{|\widehat h_{\ell,k}^{F}(t)|}.
		\label{eq:r_forward}
	\end{equation}
	The candidate value is accepted if
	\begin{equation}
		r_{\ell,k}^{F}(t)\le r_{\max},
		\label{eq:forward_acceptance}
	\end{equation}
	where $r_{\max}$ is the prescribed tolerance. If
	the inequality~\eqref{eq:forward_acceptance} is satisfied, then
	\begin{equation}
		\widehat h_{\ell,k}(t)\leftarrow\widehat h_{\ell,k}^{F}(t),
		\qquad
		E_{\ell,k}(t)\leftarrow E_{\ell,k}^{F}(t),
	\end{equation}
	and forward propagation continues. 
If Eq.~\eqref{eq:forward_acceptance} is not satisfied for at least one retained index in the monitored forward sequence, the sequence is instead reconstructed from the ratios of
consecutive terms,
\begin{equation}\label{eq:qlk_def}
	q_{\ell,k}(t):=\frac{h_{\ell,k}(t)}{h_{\ell-1,k}(t)},\qquad \ell\ge1,
\end{equation}
which follow a recurrence whose backward direction avoids this cancellation in the
region $\kappa_k(t)<0$. 
For $\ell\ge2$, dividing Eq.~\eqref{eq:rec_h} by
$h_{\ell-1,k}(t)$ gives the ratio recurrence,
\begin{equation}\label{eq:Fl_k}
	q_{\ell,k}(t)=\frac{\kappa_k(t)+\rho_k/q_{\ell-1,k}(t)}{\ell-1}.
\end{equation}
Solving for the previous ratio gives
\begin{equation}\label{eq:qbackward}
	q_{\ell-1,k}(t)=\frac{\rho_k}{(\ell-1)q_{\ell,k}(t)-\kappa_k(t)}.
\end{equation}
For $\kappa_k(t)<0$ the denominator in Eq.~\eqref{eq:qbackward} equals
\begin{equation}
	(\ell-1)q_{\ell,k}(t)+|\kappa_k(t)|,
\end{equation}
a sum of positive terms, since $q_{\ell,k}(t)>0$. The backward step is therefore
free of the subtractive cancellation present in the forward form.

Setting $\ell=m+1$ in Eq.~\eqref{eq:qbackward} expresses the ratio at index $m$
through the next one,
\begin{equation}\label{eq:qm_step}
	q_{m,k}(t)=\frac{\rho_k}{|\kappa_k(t)|+m\,q_{m+1,k}(t)}.
\end{equation}
Substituting this identity for $q_{m+1,k}$, then for $q_{m+2,k}$, and so on,
unfolds the right-hand side into the continued fraction
\begin{equation}\label{eq:qcf}
	q_{m,k}(t)
	=
	\cfrac{\rho_k}{
		|\kappa_k(t)|
		+
		\cfrac{m\rho_k}{
			|\kappa_k(t)|
			+
			\cfrac{(m+1)\rho_k}{
				|\kappa_k(t)|+\ddots}}}.
\end{equation}
As shown below, this continued fraction converges in the backward region $\kappa_k(t)<0$ and may therefore be evaluated by standard continued-fraction algorithms, such as the modified Lentz algorithm~\cite{NIST:DLMF,Lentz1976}.

Let $L_{\rm req}=L$ for density evaluation and $L_{\rm req}=L+1$ when the
full analytical Jacobian is required, because the derivative with respect
to $\theta_1$ involves $h_{\ell+1,k}$. When the backward branch is selected, any forward-generated values of the
recurrent part of the sequence are discarded.
The construction used to initialize the backward propagation is developed
below. Once $q_{L_{\rm req},k}(t)$ has been obtained,
the ratios $q_{L_{\rm req}-1,k}(t),\ldots,q_{2,k}(t)$ follow from
Eq.~\eqref{eq:qbackward}, and the recurrent part of the sequence is rebuilt
from the directly evaluated value $\widehat h_{1,k}(t)$ through
\begin{equation}
	\widehat h_{\ell,k}(t)
	=
	q_{\ell,k}(t)\widehat h_{\ell-1,k}(t),
	\qquad \ell=2,\ldots,L_{\rm req}.
\end{equation}

The backward branch is confined to the region $\kappa_k(t)<0$. For $\kappa_k(t)\ge0$ the forward recurrence combines positive contributions and is the natural low-cost propagation, whereas the backward denominator $(\ell-1)q_{\ell,k}(t)-\kappa_k(t)$ would then be a difference of positive quantities, prone to cancellation in turn. 
The forward recurrence therefore remains preferable whenever the propagated relative-error estimate does not exceed the prescribed tolerance, while the backward-ratio construction should be used in the cancellation-prone region when this condition is not satisfied.

It remains to provide a ratio $q_{N,k}(t)$ at an auxiliary index
$N\ge L_{\rm req}$, from which $q_{L_{\rm req},k}(t)$ can be obtained
by backward propagation. As shown below, the continued fraction above
provides a convergent exact representation of this ratio. Its evaluation
is iterative, however, motivating the independent large-order
approximation developed below as a low-cost algebraic alternative.
For clarity, $k$ and $t$ are fixed throughout.

For integer $\ell\ge1$, expanding the quadratic term in the Gaussian
factor of Eq.~\eqref{eq:hlk_def} and substituting $\tau=\sqrt{\nu_k}\,u$
gives
\begin{align}\label{eq:h_int_rep}
	h_{\ell,k}(t)
	&=
	C_k(t)\,
	\frac{\rho_k^{(\ell-1)/2}}{\Gamma(\ell)}
	\int_0^\infty
	u^{\ell-1}
	\exp\!\left(-\frac{u^{2}}{2}+\frac{\kappa_k(t)}{\sqrt{\rho_k}}\,u\right)du,\\
	C_k(t)&:=\frac{e^{-\frac{(t-m_k)^2}{2\nu_k}}
	}{\sqrt{2\pi}\,\theta_1}.
\end{align}
with $C_k(t)$ independent of $\ell$.  In the backward region
$\kappa_k(t)<0$, define
\begin{equation}\label{eq:X_def}
	X:=-\frac{\kappa_k(t)}{\sqrt{\rho_k}}>0 .
\end{equation}
The remaining integral can be expressed in terms of the
parabolic cylinder function $U(a,X)$, with parameter $a$
and argument $X$. Its integral representation is
\cite[Eq.~12.5.1]{NIST:DLMF}
\begin{equation}\label{eq:U_int_rep}
	U(a,X)
	=
	\frac{e^{-X^{2}/4}}{\Gamma(a+\tfrac12)}
	\int_0^\infty u^{a-1/2}e^{-u^{2}/2-Xu}\,du,
	\qquad a>-\tfrac12 .
\end{equation}
Setting $a=\ell-\tfrac12$ identifies the integral in
Eq.~\eqref{eq:h_int_rep} with
$\Gamma(\ell)\,e^{X^{2}/4}U(\ell-\tfrac12,X)$, so that
\begin{equation}\label{eq:h_pcf}
	h_{\ell,k}(t)
	=
	C_k(t)\,e^{X^{2}/4}\,\rho_k^{(\ell-1)/2}\,
	U\!\left(\ell-\tfrac12,X\right).
\end{equation}
Only the ratio of consecutive Gaussian–Gamma terms is needed. The common factor \(C_k(t)e^{X^2/4}\) cancels in this ratio: defining $R(a,X):=U(a,X)/U(a-1,X)$,
\begin{equation}\label{eq:q_pcf}
	q_{\ell,k}(t)
	=
	\sqrt{\rho_k}\;R\!\left(\ell-\tfrac12,X\right)
\end{equation}
holds exactly.  Thus, direct evaluation of the two parabolic-cylinder functions provides an exact route to the required ratio. Independently, $R(a,X)$ has the
known continued-fraction representation~\cite[Eq.~16.5.7]{Cuyt2008}
\begin{equation}\label{eq:R_cf}
	R(a,X)
	=
	\cfrac{1}{
		X+\cfrac{a+\tfrac12}{
			X+\cfrac{a+\tfrac32}{
				X+\ddots}}},
	\qquad X>0 .
\end{equation}
With $a=m-\tfrac12$, $X=|\kappa_k|/\sqrt{\rho_k}$, and
$q_{m,k}=\sqrt{\rho_k}R$, Eq.~\eqref{eq:R_cf} is precisely
Eq.~\eqref{eq:qcf}, thereby establishing the convergence asserted above.
An independent large-order approximation to the same ratio is developed
below.

Functions of consecutive order satisfy~\cite[Eq.~12.8.3]{NIST:DLMF}
$\partial_X U(a,X)-\tfrac{X}{2}U(a,X)+U(a-1,X)=0$, so that, by defining  $\Psi(a,X):=-\partial_XU(a,X)/U(a,X)$,
\begin{equation}\label{eq:R_Psi}
	R(a,X)=\frac{1}{X/2+\Psi(a,X)} .
\end{equation}
For $a>-\tfrac12$ and $X\ge0$, it follows from the integral
representation~\eqref{eq:U_int_rep} that $U>0$ and
$\partial_XU<0$, hence $\Psi>0$: the denominator in
Eq.~\eqref{eq:R_Psi} is a sum of positive quantities, free of
subtractive cancellation.  Since $U(a,X)$ satisfies the Weber equation
$\partial_X^{2}U=(a+X^{2}/4)\,U$~\cite[Eq.~12.2.2]{NIST:DLMF},
substituting $\Psi=-\partial_XU/U$ gives the first-order Riccati
equation
\begin{equation}\label{eq:riccati}
	\partial_X\Psi
	=
	\Psi^{2}-\left(a+\frac{X^{2}}{4}\right).
\end{equation}
All transformations up to this point are exact.
For large positive $a$, the terms $a$ and $X^2/4$ in
Eq.~\eqref{eq:riccati} are comparable when
$X=O(\sqrt{a})$. This motivates introducing
\begin{equation}\label{eq:scaling}
	s:=\frac{X}{2\sqrt{a}},
	\quad
	\Psi(a,X)=\sqrt{a}\,F(s;a),
	\quad
	Q(s):=\sqrt{1+s^{2}},
	\quad
	\varepsilon:=\frac{1}{2a},
\end{equation}
and using $a+X^{2}/4=aQ^{2}$ together with
$\partial_X\Psi=\tfrac12\,\partial_sF$, Eq.~\eqref{eq:riccati} becomes
\begin{equation}\label{eq:riccati_scaled}
	F^{2}-Q^{2}=\varepsilon\,\partial_sF .
\end{equation}

To construct the large-order approximation, $F$ is written formally as
a power series in $\varepsilon$:
\begin{equation}\label{eq:F_expansion}
	F(s;a)\sim\sum_{n=0}^{\infty}\varepsilon^n F_n(s),
\end{equation}
where the coefficients $F_n(s)$ contain no explicit dependence
on $\varepsilon$.
Only $F$ is expanded in Eq.~\eqref{eq:riccati_scaled}, while $Q(s)$
is kept unchanged.  Thus,
\[
\left(F_0+\varepsilon F_1+\varepsilon^2F_2+\cdots\right)^2-Q^2
=
\varepsilon\left(
\frac{dF_0}{ds}
+\varepsilon\frac{dF_1}{ds}
+\cdots
\right).
\]
The terms independent of $\varepsilon$ give
\[
F_0^2=Q^2.
\]
The zeroth-order equation admits $F_0=\pm Q$. Since the exact quantity
$F=\Psi/\sqrt{a}$ is positive, the positive leading term is retained:
\[
F_0=Q.
\]

At order $\varepsilon^n$, the two products involving $F_0=Q$ give
$2QF_n$, the remaining products give
$\sum_{j=1}^{n-1}F_jF_{n-j}$, and the right-hand side contributes
$dF_{n-1}/ds$.  Hence,
\begin{equation}\label{eq:Fn_rec}
	F_n
	=
	\frac{1}{2Q}
	\left[
	\frac{dF_{n-1}}{ds}
	-\sum_{j=1}^{n-1}F_jF_{n-j}
	\right],
	\qquad n\ge1,
\end{equation}
where the sum is absent for $n=1$.  The first two corrections are
\begin{equation}
	F_1=\frac{s}{2Q^2},
	\qquad
	F_2=\frac{2-3s^2}{8Q^5}.
\end{equation}

Inspection of further coefficients generated by
Eq.~\eqref{eq:Fn_rec} shows that their numerators are polynomials in
$s$, while their denominators follow the recurring form
$Q^{3n-1}$.  This motivates the factorization
\begin{equation}\label{eq:Fn_factorization}
	F_n(s)=\frac{P_n(s)}{Q(s)^{3n-1}},
	\qquad P_0=1,
\end{equation}
which separates the predictable denominator from the polynomial
dependence on $s$ and converts Eq.~\eqref{eq:Fn_rec} into a closed
recurrence for the polynomial numerators $P_n$.

For $n=0$, Eq.~\eqref{eq:Fn_factorization} gives
$F_0=P_0Q=Q$, since $P_0=1$. To prove the factorization by induction,
assume that Eq.~\eqref{eq:Fn_factorization} holds for
$F_0,\ldots,F_{n-1}$. Then
\[
\frac{d}{ds}
\left[
\frac{P_{n-1}}{Q^{3n-4}}
\right]
=
\frac{
	(1+s^2)\dfrac{dP_{n-1}}{ds}
	-(3n-4)sP_{n-1}
}{
	Q^{3n-2}
},
\]
where $dQ/ds=s/Q$, while
\[
F_jF_{n-j}
=
\frac{P_jP_{n-j}}{Q^{3n-2}}.
\]
All terms inside the brackets of Eq.~\eqref{eq:Fn_rec} therefore have
the same denominator $Q^{3n-2}$, and the additional factor $1/(2Q)$
produces the required denominator $Q^{3n-1}$.  The resulting numerator
satisfies
\begin{equation}\label{eq:Pn_rec}
	P_n
	=
	\frac12
	\left[
	(1+s^2)\frac{dP_{n-1}}{ds}
	-(3n-4)sP_{n-1}
	-\sum_{j=1}^{n-1}P_jP_{n-j}
	\right],
	\qquad n\ge1.
\end{equation}

Equation~\eqref{eq:Pn_rec} recursively generates polynomials $P_n(s)$ of degree at most $n$.  Moreover, $P_n$ is even for even $n$ and odd for odd $n$,
\[
P_n(-s)=(-1)^nP_n(s),
\]
 so it contains only powers of $s$ with the same parity as $n$. Moreover, the coefficients of the $P_n$ are independent of $a$ and of all physical parameters. Consequently, for any chosen truncation order $p$, the polynomials $P_1,\ldots,P_p$ need to be generated from Eq.~\eqref{eq:Pn_rec} only once and can subsequently be reused for all evaluations. Thus, the expansion becomes 
\begin{equation}\label{eq:F_ansatz}
	F(s;a)
	\sim
	\sum_{n=0}^{\infty}
	\varepsilon^n
	\frac{P_n(s)}{Q(s)^{3n-1}},
	\qquad P_0=1.
\end{equation}

For a truncation after $p$ corrections, factoring out the leading
factor $Q$ gives
\begin{equation}
	F^{(p)}(s;a)
	=
	Q(s)\,H_p(s;a),
\end{equation}
where
\begin{equation}\label{eq:Hp_def}
	H_p(s;a)
	:=
	1+\sum_{n=1}^{p}b^nP_n(s),
	\qquad
	b:=\frac{\varepsilon}{Q^3}
	=\frac{1}{2aQ^3}.
\end{equation}
The finite sum can be written in nested form as
\begin{equation}
	H_p(s;a)
	=
	1+b\bigl[P_1+b\bigl(P_2+\cdots+bP_p\bigr)\bigr].
\end{equation}

Combining $F\approx F^{(p)}=QH_p$ with Eqs.~\eqref{eq:q_pcf}
and~\eqref{eq:R_Psi}, the ratio at an auxiliary index $N$,
with $a=N-\tfrac12$, is approximated by
\begin{equation}\label{eq:qN_terminal}
	q^{(p)}_{N,k}(t)
	=
	\frac{\sqrt{\rho_k}}
	{X/2+\sqrt{a}\,Q(s)\,H_p(s;a)},
	\qquad
	s=\frac{X}{2\sqrt{a}}.
\end{equation}
This expression involves only elementary operations and requires neither
$h_{N,k}$ nor $U(a,X)$ separately.

Since the degree and algebraic complexity of the correction polynomials
increase with $p$, it is advantageous to retain a moderate truncation
order and to increase instead the auxiliary index $N$ when a more accurate
terminal-ratio approximation is required. Numerical exploration shows that the
terminal-ratio approximation becomes progressively more demanding as $X$ is
reduced over the region relevant to backward propagation. For $p=10$,
increasing the auxiliary index to $N=30$ was found to bring the
terminal-ratio error close to the level of double-precision roundoff even
in the more demanding small-$X$ region. At $X=0$, a slightly smaller terminal-ratio error was observed.  At $s=0$ the parity relation $P_n(-s)=(-1)^nP_n(s)$ implies that all odd-order polynomials vanish.
For $p=10$, the first omitted odd-order polynomial, $P_{11}$, therefore
vanishes at $s=0$.
Moreover, extremely small positive $X$ produces little cancellation
in the forward recurrence.  Indeed, using
$q_{\ell-1,k}=\sqrt{\rho_k}\,R(\ell-\tfrac32,X)$, the magnitude of the
subtracted term in the forward recurrence relative to the positive one is
\[
\frac{|\kappa_k(t)|h_{\ell-1,k}(t)}
{\rho_k h_{\ell-2,k}(t)}
=
X\,R\!\left(\ell-\frac32,X\right).
\]
This quantity vanishes as $X\to0^+$.  Thus, the cancellation that motivates backward propagation disappears in this limit, favouring forward evaluation when the propagated relative-error estimate remains within the prescribed tolerance.
Different balances between the truncation order $p$ and the auxiliary index $N$ are therefore possible. One convenient low-order choice is $p=10$ together with
\begin{equation}
	N=\max\{30,L_{\rm req}\}.
\end{equation}

If $N>L_{\rm req}$, the required ratio $q_{L_{\rm req},k}(t)$ may be
obtained by applying the backward recurrence~\eqref{eq:qbackward} from
$N$ to $L_{\rm req}$. Since $\{U(a+n,X)\}_{n\ge0}$ is a minimal
solution of the corresponding three-term recurrence for
$X>0$~\cite{Cuyt2008}, backward propagation is the numerically stable
direction for this solution~\cite{Gautschi1967}. Only
$N-L_{\rm req}$ additional scalar recurrence steps are required. Consequently, little practical benefit would be gained from a more elaborate $X$-dependent optimization of $N$, or from dynamically varying the truncation order $p$, since the additional backward steps are inexpensive and a fixed low-order construction such as the one above already yields terminal-ratio errors close to double-precision roundoff, well below those relevant for the intended chromatographic calculations.

\section{Analytical derivatives}\label{sec:derivatives}

The Jacobian of $f_T^{(L,n)}$ is assembled from derivatives of its constituent blocks, obtained below.

\subsection{Slow-retention coefficients}
In Ref.~\cite{Sanchez2026b}, the parameter derivatives of $\Omega_\ell$ were obtained by differentiating the original coefficient
recurrence, yielding an exact secondary recurrence with quadratic scaling in the retained order $L$.  The present linear-scaling formulation instead generates the auxiliary quantities $C_{j,\ell}$ and $D_{j,\ell}$ as part of the coefficient evaluation.  The derivatives are rederived below as closed expressions in terms of these same quantities, requiring only a few additional products and differences, together with a single sum over the mechanisms at each order for the derivative with respect to $\theta_1$. Consequently, no separate derivative recurrence is required, and all
$2M$ parameter derivatives can be obtained with low additional computational cost while preserving the $\mathcal{O}(LM)$ scaling of the coefficient calculation.

Let
\begin{equation}\label{eq:G_def}
	G(u):=\sum_{\ell=0}^{\infty}\Omega_\ell u^\ell
\end{equation}
denote the generating function of the coefficients.  Since
$p_\ell=e^{-\Lambda}\Omega_\ell$, this function is related to the
probability generating function in Eq.~\eqref{eq:GK_explicit} by
\[
G(u)=e^\Lambda G_K(u).
\]
Substitution of Eq.~\eqref{eq:GK_explicit} therefore gives
\begin{equation}\label{eq:Omega_generating_function}
	G(u)
	=
	\exp\left\{
	\sum_{j=1}^{M}
	\Lambda_j
	\frac{(1-a_j)u}{1-a_j u}
	\right\}.
\end{equation}
By definition, $\Omega_\ell$ is the coefficient multiplying $u^\ell$ in
the power-series expansion of $G(u)$.  Since $u$ is an auxiliary variable
independent of the model parameters, differentiation with respect to any
parameter $p$ acts only on the coefficients:
\begin{equation}
	\frac{\partial G(u)}{\partial p}
	=
	\sum_{\ell=0}^{\infty}
	\frac{\partial\Omega_\ell}{\partial p}\,u^\ell.
\end{equation}
Consequently, $\partial\Omega_\ell/\partial p$ equals the coefficient of $u^\ell$ once $\partial G(u)/\partial p$ has been expressed as a
power series.

By the chain rule,
\begin{equation}\label{eq:dG_dLambda}
	\frac{\partial G}{\partial\Lambda_j}
	=
	G(u)\,
	\frac{(1-a_j)u}{1-a_j u}.
\end{equation}
Expanding the quotient as
\[
\frac{u}{1-a_j u}
=
\sum_{r\ge1}a_j^{r-1}u^r
\]
and multiplying by
\[
G(u)=\sum_{s\ge0}\Omega_s u^s,
\]
the coefficient of $u^\ell$ in the product collects the terms whose
orders satisfy $r+s=\ell$.  Since $r\ge1$ and $s\ge0$, these contributions correspond to $r=1,\ldots,\ell$, with $s=\ell-r$. The resulting coefficient is therefore
\begin{equation}
	\sum_{r=1}^{\ell}
	a_j^{r-1}\Omega_{\ell-r}
	=
	D_{j,\ell},
\end{equation}
which is precisely the auxiliary sum defined in Eq.~\eqref{eq:D_def}.
It follows that, for $\ell\ge1$,
\begin{equation}\label{eq:dOmega_dLambda_CD}
	\frac{\partial\Omega_\ell}{\partial\Lambda_j}
	=
	(1-a_j)D_{j,\ell}.
\end{equation}

To obtain the scale derivatives, first differentiate $G(u)$ with respect to the auxiliary parameter $a_j$:
\begin{equation}\label{eq:dG_da}
	\frac{\partial G}{\partial a_j}
	=
	\Lambda_j
	G(u)
	\frac{u(u-1)}{(1-a_j u)^2}.
\end{equation}
Because
\begin{equation}
	\frac{u}{(1-a_j u)^2}
	=
	\sum_{r\ge1}
	r a_j^{r-1}u^r,
\end{equation}
after multiplication by $G(u)$, the coefficient of $u^\ell$ in
$G(u)u/(1-a_j u)^2$ is
\begin{equation}
	\sum_{r=1}^{\ell}
	r a_j^{r-1}\Omega_{\ell-r}
	=
	C_{j,\ell},
\end{equation}
which is the auxiliary sum in Eq.~\eqref{eq:Omega_C_rec}. 
Thus,
\[
G(u)\frac{u}{(1-a_j u)^2}
=
\sum_{\ell\ge1}C_{j,\ell}u^\ell.
\]
Multiplication by $u$ raises the power of every term by one, so the coefficient at order $u^\ell$ changes from $C_{j,\ell}$ to $C_{j,\ell-1}$, with $C_{j,0}=0$ as implied by the empty sum in Eq.~\eqref{eq:Omega_C_rec}. Thus, multiplying this series by $u-1$ gives
\[
(u-1)\sum_{\ell\ge1}C_{j,\ell}u^\ell
=
\sum_{\ell\ge1}
\left(
C_{j,\ell-1}-C_{j,\ell}
\right)u^\ell.
\]
Consequently,
Eq.~\eqref{eq:dG_da} can be written as
\[
\frac{\partial G}{\partial a_j}
=
\Lambda_j
\sum_{\ell\ge1}
\left(
C_{j,\ell-1}-C_{j,\ell}
\right)u^\ell.
\]
Matching the coefficient of $u^\ell$ with the termwise derivative of
$G(u)$ therefore gives, for $\ell\ge1$,
\begin{equation}\label{eq:dOmega_da_CD}
	\frac{\partial\Omega_\ell}{\partial a_j}
	=
	\Lambda_j
	\left(
	C_{j,\ell-1}-C_{j,\ell}
	\right).
\end{equation}

The scale derivatives follow from Eq.~\eqref{eq:dOmega_da_CD} by the
chain rule. Since $a_1\equiv0$, $\partial a_1/\partial\theta_1=0$,
whereas for $j\ge2$,
$\partial a_j/\partial\theta_1=-1/\theta_j$ and
$\partial a_j/\partial\theta_j=\theta_1/\theta_j^2$.
For the reference scale,
\begin{equation}	\label{eq:dOmega_dtheta1_CD}
	\frac{\partial\Omega_\ell}{\partial\theta_1}
	=
	\sum_{j=2}^{M}
	\frac{\partial\Omega_\ell}{\partial a_j}
	\frac{\partial a_j}{\partial\theta_1}
	=
	-\sum_{j=2}^{M}
	\frac{\Lambda_j}{\theta_j}
	\left(
	C_{j,\ell-1}-C_{j,\ell}
	\right),
	\qquad \ell\ge1.
\end{equation}
For $j\ge2$,
\begin{equation}
	\frac{\partial\Omega_\ell}{\partial\theta_j}
	=
	\Lambda_j
	\frac{\theta_1}{\theta_j^2}
	\left(
	C_{j,\ell-1}-C_{j,\ell}
	\right),
	\qquad \ell\ge1.
	\label{eq:dOmega_dthetaj_CD}
\end{equation}
All derivatives vanish at $\ell=0$, since $\Omega_0=1$. 
These expressions apply as long as $\theta_1$ remains the unique minimum
scale.  As discussed above, mechanisms with identical scales can be merged exactly, so this degeneracy does not pose a practical difficulty in the adopted parametrization.

\subsection{Gaussian--Gamma derivatives}

The functions $h_{\ell,k}(t)=h_\ell(t;m_k,\nu_k,\theta_1)$ are instances
of the Gaussian--Gamma family derived in Ref.~\cite{Sanchez2026b}; the
derivatives with respect to their arguments obtained there apply
directly.  For $\ell\ge1$,
\begin{equation}\label{eq:dh_dm}
	\frac{\partial h_{\ell,k}}{\partial m_k}
	=
	\frac{1}{\theta_1}
	\left(
	h_{\ell,k}-h_{\ell-1,k}
	\right),
\end{equation}
while
\begin{equation}\label{eq:dh0_dm}
	\frac{\partial h_{0,k}}{\partial m_k}
	=
	\frac{t-m_k}{\nu_k}h_{0,k}.
\end{equation}
For $\ell\ge2$,
\begin{equation}\label{eq:dh_dv}
	\frac{\partial h_{\ell,k}}{\partial \nu_k}
	=
	\frac{1}{2\theta_1^2}
	\left(
	h_{\ell-2,k}
	-
	2h_{\ell-1,k}
	+
	h_{\ell,k}
	\right),
\end{equation}
with the boundary cases
\begin{equation}\label{eq:dh0_dv}
	\frac{\partial h_{0,k}}{\partial \nu_k}
	=
	\left[
	\frac{(t-m_k)^2}{2\nu_k^2}
	-
	\frac{1}{2\nu_k}
	\right]h_{0,k},
\end{equation}
and
\begin{equation}\label{eq:dh1_dv}
	\frac{\partial h_{1,k}}{\partial \nu_k}
	=
	\frac{1}{2\theta_1^2}
	\left(
	h_{1,k}-h_{0,k}
	\right)
	-
	\frac{t-m_k}{2\theta_1\nu_k}h_{0,k}.
\end{equation}
For the common Gamma scale,
\begin{equation}\label{eq:dh_dtheta1}
	\frac{\partial h_{\ell,k}}{\partial\theta_1}
	=
	\frac{\ell}{\theta_1}
	\left(
	h_{\ell+1,k}-h_{\ell,k}
	\right),
	\qquad
	\ell\ge1,
\end{equation}
and $\partial h_{0,k}/\partial\theta_1=0$.  Evaluating the $\theta_1$-derivative at truncation order $L$ therefore
requires $h_{\ell,k}$ through $\ell=L+1$, one order beyond the maximum retained index.

\subsection{Physical-parameter derivatives of the quadrature components}

The quantities entering Eq.~\eqref{eq:fT_master} through the quadrature
components are $W_k$, $\nu_k$, and $m_k$.  With the column length $x$
and mobile-phase velocity $v$ held fixed, Eqs.~\eqref{eq:NIG_phys_map},
\eqref{eq:nuk_Wk_mk}, and~\eqref{eq:inj_conv} give
\begin{equation}\label{eq:physical_quadrature_reduced}
	W_k
	=
	\frac{2\omega_k}{1+u_k},
	\qquad
	\nu_k
	=
	\frac{x\sigma_f^2}{v}\,u_k+\sigma_{\mathrm{inj}}^2,
	\qquad
	m_k
	=
	\frac{x(1+\mu_f)}{v}\,u_k .
\end{equation}

The remaining implicit parameter dependence is contained in $u_k$.
From Eq.~\eqref{eq:NIG_phys_map}, together with
$\gamma=\sqrt{\alpha^2-\beta^2}$ and $\sigma=\sqrt{\gamma\delta}$,
\[
\sigma=\sqrt{\frac{vx}{2D_{\mathrm{eff}}}}.
\]
Thus, among the parameters considered here, $u_k$ depends only on
$D_{\mathrm{eff}}$.  Differentiation of Eq.~\eqref{eq:uk} then gives
\begin{equation}\label{eq:duk_dDeff}
	\frac{\partial u_k}{\partial D_{\mathrm{eff}}}
	=
	\frac{z_k u_k}
	{2D_{\mathrm{eff}}\sigma
		\sqrt{1+z_k^2/(4\sigma^2)}} .
\end{equation}

The nonzero derivatives of the quadrature quantities are therefore
\begin{equation}\label{eq:dWk}
	\frac{\partial W_k}{\partial D_{\mathrm{eff}}}
	=
	-\frac{2\omega_k}{(1+u_k)^2}
	\frac{\partial u_k}{\partial D_{\mathrm{eff}}},
\end{equation}
\begin{equation}\label{eq:dvk}
	\frac{\partial\nu_k}{\partial D_{\mathrm{eff}}}
	=
	\frac{x\sigma_f^2}{v}
	\frac{\partial u_k}{\partial D_{\mathrm{eff}}},
	\qquad
	\frac{\partial\nu_k}{\partial\sigma_f^2}
	=
	\frac{x}{v}\,u_k,
	\qquad
	\frac{\partial\nu_k}{\partial\sigma_{\mathrm{inj}}^2}
	=
	1,
\end{equation}
and
\begin{equation}\label{eq:dmk}
	\frac{\partial m_k}{\partial D_{\mathrm{eff}}}
	=
	\frac{x(1+\mu_f)}{v}
	\frac{\partial u_k}{\partial D_{\mathrm{eff}}},
	\qquad
	\frac{\partial m_k}{\partial\mu_f}
	=
	\frac{x}{v}\,u_k .
\end{equation}
All other derivatives of $W_k$, $\nu_k$, and $m_k$ with respect to
$D_{\mathrm{eff}}$, $\mu_f$, $\sigma_f^2$, and
$\sigma_{\mathrm{inj}}^2$ vanish.

\subsection{Full Jacobian assembly}

To abbreviate the inner node sum of Eq.~\eqref{eq:fT_master}, which
recurs in the derivatives with respect to the slow-retention
parameters, define
\begin{equation}\label{eq:Aell_def}
	A_\ell(t)
	:=
	\sum_{k=1}^{n}
	W_k h_{\ell,k}(t).
\end{equation}
These sums may be accumulated during the density evaluation and reused
in the derivatives below.

For $\Lambda_j$,
\begin{equation}\label{eq:df_dLambda}
	\frac{\partial f_T^{(L,n)}}{\partial\Lambda_j}
	=
	e^{-\Lambda}
	\sum_{\ell=0}^{L}
	\left(
	\frac{\partial\Omega_\ell}{\partial\Lambda_j}
	-
	\Omega_\ell
	\right)
	A_\ell(t).
\end{equation}
For a non-reference slow-retention scale $\theta_j$, with $j\ge2$,
\begin{equation}\label{eq:df_dthetaj}
	\frac{\partial f_T^{(L,n)}}{\partial\theta_j}
	=
	e^{-\Lambda}
	\sum_{\ell=0}^{L}
	\frac{\partial\Omega_\ell}{\partial\theta_j}
	A_\ell(t).
\end{equation}
For the reference scale $\theta_1$, both the slow-retention
coefficients and the Gaussian--Gamma functions contribute:
\begin{equation}\label{eq:df_dtheta1}
	\begin{aligned}
		\frac{\partial f_T^{(L,n)}}{\partial\theta_1}
		={}&
		e^{-\Lambda}
		\sum_{\ell=0}^{L}
		\frac{\partial\Omega_\ell}{\partial\theta_1}
		A_\ell(t)
		\\
		&+
		e^{-\Lambda}
		\sum_{\ell=1}^{L}
		\Omega_\ell
		\frac{\ell}{\theta_1}
		\left[
		A_{\ell+1}(t)-A_\ell(t)
		\right],
	\end{aligned}
\end{equation}
which requires $A_\ell(t)$ up to $\ell=L+1$.

For a mobile-plus-fast parameter $p$, which does not affect the
slow-retention parameters, define
\begin{equation}\label{eq:dAell_dp}
	\begin{aligned}
		\partial_p A_\ell(t)
		=
		\sum_{k=1}^{n}
		\bigg[
		&
		\partial_p W_k\,h_{\ell,k}(t)
		\\
		&+
		W_k
		\left(
		\frac{\partial h_{\ell,k}}{\partial m_k}
		\partial_p m_k
		+
		\frac{\partial h_{\ell,k}}{\partial \nu_k}
		\partial_p \nu_k
		\right)
		\bigg].
	\end{aligned}
\end{equation}
The corresponding density derivative is
\begin{equation}\label{eq:df_dp}
	\partial_p f_T^{(L,n)}(t)
	=
	e^{-\Lambda}
	\sum_{\ell=0}^{L}
	\Omega_\ell
	\partial_p A_\ell(t).
\end{equation}
The same values of $h_{\ell,k}(t)$ generated for the density evaluation
are reused in Eq.~\eqref{eq:dAell_dp}; no additional Gaussian--Gamma
convolution evaluations are required.

\section{Independent numerical verification}\label{sec:validation}

The time-domain density and its analytical derivatives, evaluated in double precision (IEEE 754 binary64) using Fortran, were compared with high-precision references computed in Python by an independent route: numerical inversion of the corresponding characteristic function and its analytical derivatives. The double-precision evaluation of the Gaussian--Gamma recurrence was assessed separately against high-precision reference sequences computed by forward recurrence using Python.

A synthetic multimechanism test case was constructed for numerical verification, with its transport and fast-retention parameters based on the reported glucose/H$^+$ system of Ref.~\cite{Wisniewski2014}:
\begin{equation}
	\begin{aligned}
		x&=0.110\ {\rm m},
		&v&\simeq4.953\times10^{-4}\ {\rm m\,s^{-1}},\\
		D_{\mathrm{eff}}&\simeq2.179\times10^{-7}\ {\rm m^2\,s^{-1}},
		&\mu_f&\simeq0.8437,\\
		\sigma_f^2&\simeq3.437\ {\rm s}
	\end{aligned}
\end{equation}
The column-scale P\'eclet number is $\mathrm{Pe}=250$.
The variance of the Gaussian entry-time distribution was matched to that of a uniform entry-time distribution corresponding to a rectangular injection pulse of duration $V_{\mathrm{inj}}/Q$, using the injection volume $V_{\mathrm{inj}}=100\ \mu{\rm L}$ and volumetric flow rate $Q=4.97\ {\rm mL\,min^{-1}}$ reported in Ref.~\cite{Wisniewski2014}:
\begin{equation}
\sigma_{\mathrm{inj}}^2=(V_{\mathrm{inj}}/Q)^2/12
\simeq 0.1215\ {\rm s^2}.
\end{equation}
Three synthetic slow mechanisms were introduced solely for numerical
validation of the multimechanism formulation:
\begin{equation}
	(\Lambda_1,\Lambda_2,\Lambda_3)=(0.6,0.3,0.1),
	\qquad
	(\theta_1,\theta_2,\theta_3)=(10,30,60)\ {\rm s}.
\end{equation}
Except for the NIG--Gaussian comparison described below, all comparisons used the 41 times $t_i=(250+15i)\ {\rm s}$, $i=0,\ldots,40$. Unless otherwise specified, $n=100$, $L=253$, and $r_{\max}=10^{-10}$ were used.
The recurrence was evaluated through
$L_{\rm req}=L+1=254$, with density sums and component statistics
restricted to $0\le\ell\le L$.

\subsection{Complete density and quadrature convergence}
The independent reference density was obtained by Fourier inversion of the characteristic function of the total elution time. This function follows directly from the independent-sum representation in Eq.~\eqref{eq:Tdecomp}, using the NIG and Gaussian characteristic functions and the standard compound-Poisson formula:
\begin{equation}\label{eq:CF_validation}
	\begin{aligned}
		\varphi_T(u)&=\varphi_{\mathrm{NIG}}(u)
		\exp\!\left(-\frac{\sigma_{\mathrm{inj}}^2u^2}{2}\right)\\
		&\quad\times\prod_{j=1}^{M}
		\exp\!\left\{\Lambda_j\left[(1-iu\theta_j)^{-1}-1\right]\right\},
	\end{aligned}
\end{equation}
where
\begin{equation}\label{eq:CF_NIG_validation}
	\varphi_{\mathrm{NIG}}(u)=
	\exp\!\left[i\mu u+\delta\left(
	\gamma-\sqrt{\alpha^2-(\beta+iu)^2}\right)\right],
\end{equation}
with the parameters given by Eq.~\eqref{eq:NIG_phys_map}.
The inversion was performed at 80 decimal digits over
$0\le u\le0.5\ {\rm s^{-1}}$, using 256-point Gauss--Legendre
quadrature on panels of width $0.05\ {\rm s^{-1}}$.
A comparison calculation at 60 digits, with cutoff
$0.4\ {\rm s^{-1}}$ and 128 points per panel, gave the same
binary64-rounded density and Jacobian values at every time.
These rounded references were used to form the density and Jacobian
discrepancies reported below.

For the slow-series truncation, high-precision evaluation gives
\begin{equation}\label{eq:RL_validation}
	R_L=1-e^{-\Lambda}\sum_{\ell=0}^{L}\Omega_\ell,
	\qquad R_{253}=8.76879\times10^{-21}.
\end{equation}
For the default quadrature order $n=100$, the maximum density discrepancy, normalized by the reference peak
height on the grid, was
\begin{equation}\label{eq:Einf_validation}
	E_\infty=
	\frac{\max_i|f_T^{\rm DP}(t_i)-f_T^{\rm CF}(t_i)|}
	{\max_i f_T^{\rm CF}(t_i)}
	=1.36\times10^{-15}.
\end{equation}
Here the superscripts $\mathrm{DP}$ and $\mathrm{CF}$ denote, respectively, the double-precision time-domain values and the high-precision characteristic-function-inversion references. 
The maximum pointwise relative discrepancy,
$\max_i\!\left[|f_T^{\rm DP}(t_i)-f_T^{\rm CF}(t_i)|/
f_T^{\rm CF}(t_i)\right]$,
was $3.94\times10^{-15}$.
Quadrature convergence was examined by varying $n$ at fixed physical
parameters, $L$, and $r_{\max}$ (Table~\ref{tab:quadrature_validation}).
The decrease in the discrepancy is followed by fluctuations of order
$10^{-15}$ in the double-precision evaluation.

\begin{table}[H]
	\centering
	\caption{Maximum normalized density discrepancy $E_\infty$ (Eq.~\eqref{eq:Einf_validation}) as a function of the number of NIG quadrature nodes $n$, with $L=253$ and $r_{\max}=10^{-14}$.}
	\label{tab:quadrature_validation}
\begin{tabular}{@{}r@{\quad}c@{\qquad}r@{\quad}c@{}}
	\toprule
	$n$ & $E_\infty$ & $n$ & $E_\infty$ \\
	\midrule
		20 & $3.32\times10^{-7}$  & 60  & $5.68\times10^{-16}$ \\
		30 & $3.52\times10^{-10}$ & 70  & $1.48\times10^{-15}$ \\
		40 & $2.37\times10^{-13}$ & 80  & $7.95\times10^{-16}$ \\
		50 & $1.14\times10^{-15}$ & 100 & $1.36\times10^{-15}$ \\
		\bottomrule
	\end{tabular}
\end{table}

As an illustrative comparison for the validation case, the density was
also evaluated with a Gaussian mobile-plus-fast contribution having the
same mean and variance, while retaining the same slow-retention and
injection contributions. On a grid from 50 to 1100 s at 0.5 s intervals,
the maximum absolute difference between the two densities was $4.34\%$ of the NIG-base peak height on that grid.

\subsection{Gaussian--Gamma recurrence}

The high-precision (HP) reference values $h_{\ell,k}^{\mathrm {HP}}$ were calculated using only the forward recurrence from the exact binary64 inputs $m_k$, $\nu_k$, $t$, and $\theta_1$.
Working precision was increased in increments
of 50 digits, starting with comparisons at 100 and 150 digits, until all computed values through order $L+1$ were positive and, at every order, the relative difference between the values obtained at successive working precisions was less than $10^{-30}$.
The retained working precisions ranged from 150 to 400 digits.
The component errors below were calculated relative to the unrounded
HP values. The algebraic terminal-ratio approximation with $p=10$ and $N=\max(30,L_{\mathrm {req}})$ was used unless stated otherwise.

\subsubsection{Forward error indicator and switching}

For each $(t,k)$ sequence, the recurrence was also evaluated without
switching. The estimated and reference relative errors of the forward recurrence were defined as
\begin{equation}\label{eq:forward_error_validation}
	r_{\ell,k}^{F}=\frac{E_{\ell,k}^{F}}{|h_{\ell,k}^{F}|},
	\qquad
	e_{\ell,k}^{F}=\frac{|h_{\ell,k}^{F}-h_{\ell,k}^{\mathrm {HP}}|}
	{h_{\ell,k}^{\mathrm {HP}}}.
\end{equation}
For $r_{\max}=10^{-6},10^{-9},10^{-12}$ and $10^{-14}$,  the ratio
$e_{\ell,k}^{F}/r_{\ell,k}^{F}$ was evaluated for every monitored
recurrence term with $\ell\ge2$ that was accepted during forward
propagation, up to the switching order when switching occurred, or through
the complete monitored sequence otherwise
(Table~\ref{tab:switching_validation}).
\begin{table*}[t]
	\centering
	\caption{Forward-error assessment against high-precision references. Statistics of $e^F/r^F$ [Eq.~\eqref{eq:forward_error_validation}] include only monitored terms accepted at each tolerance $r_{\max}$. “Sequences switched” counts the $(t,k)$ sequences reconstructed backward. P99 denotes the 99th percentile.}
	\label{tab:switching_validation}
	\begin{tabular}{ccccccc}
		\toprule
		$r_{\max}$ &
		Sequences switched &
		Accepted monitored terms &
		\multicolumn{3}{c}{$e^F/r^F$} &
		$\displaystyle\frac{\max_{\mathrm {accepted}}e^F}{r_{\max}}
		$
		\\
		\cmidrule(lr){4-6}
		& & & Median & P99 & Maximum & \\
		\midrule
		$10^{-6}$  & 1898 & 35966 & 0.0702 & 0.411 & 1.775 & 0.602 \\
		$10^{-9}$  & 1918 & 24055 & 0.0594 & 0.400 & 1.775 & 0.547 \\
		$10^{-12}$ & 1936 & 11999 & 0.0515 & 0.400 & 1.775 & 0.524 \\
		$10^{-14}$ & 1957 &  1243 & 0.0887 & 0.659 & 1.775 & 0.400 \\
		\bottomrule
	\end{tabular}
\end{table*}
For all four tolerances, no monitored recurrence term accepted during
forward propagation had a reference relative error exceeding
$r_{\max}$. The median ratio $e^F/r^F$ ranged from 0.052 to 0.089,
while its 99th percentile did not exceed 0.659. The propagated estimate
is not a strict upper bound, since the largest observed value of
$e^F/r^F$ was 1.78, but the largest reference relative error among
accepted terms remained below $0.602\,r_{\max}$. At the first rejected
order of every sequence for which switching occurred, the propagated
estimate exceeded the reference error. Reducing $r_{\max}$ further provides little practical benefit: at $r_{\max}=10^{-15}$ the indicator begins to probe relative-error changes on the scale of accumulated double-precision roundoff, while the attainable
accuracy of the final hybrid evaluation remains essentially unchanged.

\subsubsection{Double-precision accuracy of the hybrid recurrence}

The accuracy of the hybrid recurrence was assessed by comparing the final
computed values, denoted by $h_{\ell,k}^{H}$, with the corresponding
high-precision reference values $h_{\ell,k}^{\rm HP}$. Calculations used
$r_{\max}=10^{-14}$. For the $1,039,222$ components whose reference
values lie in the normal range of double-precision arithmetic, the median,
99th-percentile, and maximum relative errors
$|h_{\ell,k}^{H}-h_{\ell,k}^{\rm HP}|/h_{\ell,k}^{\rm HP}$ were
$4.18\times10^{-15}$, $5.74\times10^{-14}$, and
$2.42\times10^{-13}$, respectively. No negative or nonfinite values were obtained.

Of the remaining components, $1145$ had subnormal reference values and $1033$ had reference values below the smallest positive representable double-precision number. Across these components, the maximum relative error was $1.00$, or only $0.858$ when restricted to the subnormal-reference values. Near the smallest positive representable double-precision number, the difference between consecutive representable numbers can be comparable to the values themselves, so relative errors of order unity may occur even for an optimally rounded double-precision result.

\subsection{Comparison of terminal-ratio evaluations}

When the continued fraction for $q_{L_{\rm req},k}$ was evaluated using Lentz's method in double precision, the iteration was stopped when $|\Delta_j-1|\le10^{-15}$, where $\Delta_j$ is the multiplicative update of the continued-fraction approximant at iteration $j$.

The algebraic approximation to the terminal ratio avoids this iterative evaluation altogether, which is particularly advantageous for small $X=|\kappa_k|/\sqrt{\rho_k}$, where direct evaluation by Lentz's method requires a large number of iterations. At the adopted $r_{\max}=10^{-14}$, the two terminal-ratio evaluations gave similar median and 99th-percentile component errors: $4.18\times10^{-15}$ and $5.74\times10^{-14}$ for the algebraic evaluation, and $4.20\times10^{-15}$ and $5.81\times10^{-14}$ for Lentz's method, respectively. The maximum relative errors were $2.42\times10^{-13}$ and $4.70\times10^{-12}$, respectively.

The largest error with Lentz's method occurred at $t=265\ {\rm s}$, $k=29$, and $\ell=252$, for $X\simeq0.00702$. Thus, the algebraic approximation is particularly advantageous for small $X$, where Lentz's method requires a large number of iterations and yielded a larger maximum error in the numerical comparison.

\subsection{Independent validation of the analytical Jacobian}

The double-precision values of the analytical derivatives of the finite time-domain representation were compared with reference values obtained independently from the characteristic-function representation.
The derivatives were considered with respect to
$\Lambda_1,\ldots,\Lambda_M$, $\theta_1,\ldots,\theta_M$, $D_{\mathrm{eff}}$, $\mu_f$ and $\sigma_f^2$, with $x$, $v$, and $\sigma_{\mathrm{inj}}^2$ fixed.  For each parameter $p$, the characteristic function in Eq.~\eqref{eq:CF_validation} was differentiated analytically. Using $\xi$
as a dummy Fourier variable,
\begin{equation}
	\left.
	\frac{\partial\varphi_T(\xi)}{\partial p}
	\right|_{\xi=u}
	=
	\varphi_T(u)
	\left.
	\frac{\partial\log\varphi_T(\xi)}{\partial p}
	\right|_{\xi=u}.
\end{equation}
The resulting derivative was then inverted using the same high-precision
Fourier quadrature employed for the density reference. This provides
reference derivatives without using the time-domain derivative expressions
or their recurrence-based evaluation.

Let $J_p^{\rm DP}$ denote the double-precision time-domain derivative,
\begin{equation}
	J_p^{\rm DP}
	:=
	\frac{\partial f_T^{(L,n)}}{\partial p}.
\end{equation}
The independently inverted reference is denoted by \(J_p^{\rm CF}(t)\). For each parameter, the discrepancy was normalized by the maximum absolute reference derivative over the same validation grid used above:

\begin{equation}\label{eq:Jac_relative_error}
	E_p=
	\frac{\max_i|J_p^{\rm DP}(t_i)-J_p^{\rm CF}(t_i)|}
	{\max_i|J_p^{\rm CF}(t_i)|}.
\end{equation}
For $n=100$, $L=253$, and $r_{\max}=10^{-10}$, the values of $E_p$
over the nine parameters ranged from $1.71\times10^{-15}$ to
$7.43\times10^{-15}$, with the largest value obtained for $\theta_1$.

\section{Conclusions}
The time-domain formulation of Refs.~\cite{Sanchez2026,Sanchez2026b} has been extended beyond the high-Péclet Gaussian approximation. The NIG mobile-plus-fast contribution is approximated by a finite Gaussian mixture, leading to an analytic time-domain expression for the elution-time PDF as a weighted sum of single-index Gaussian–Gamma series. Each series has the same form as the previously derived multi-mechanism series, with parameters specific to the corresponding mixture component. Analytical derivatives of this density with respect to the model parameters were also obtained.

The normalized series coefficients have been identified as the probability mass function of a sum of independent Pólya--Aeppli random variables. 
Using this characterization together with physical considerations, the required truncation order is found to be low enough to permit rapid evaluation of the series under usual chromatographic conditions.
New recurrence relations were derived for evaluating these coefficients and their derivatives, reducing the computational cost from quadratic to linear in the truncation order for a fixed number of mechanisms.

A backward reconstruction procedure is developed to overcome the loss of relative accuracy caused by subtractive cancellation in the forward evaluation of Gaussian--Gamma sequences. A roundoff-error estimator is also derived to quantify this loss and determine when backward reconstruction is necessary. An exact expression for the terminal ratio in terms of parabolic-cylinder functions was derived, together with a highly accurate, non-iterative algebraic approximation for its efficient evaluation.

		\section*{Acknowledgement}
		This work was supported by the National Scientific and Technical Research Council of Argentina (CONICET).
		
		\printbibliography
	\end{multicols}
\end{document}